\newcommand{\CLASSINPUTtoptextmargin}{0.76in}
\newcommand{\CLASSINPUTbottomtextmargin}{1.06in}
\documentclass[conference]{IEEEtran}

\usepackage{epsfig,rotating,setspace,latexsym,amsmath,epsf,amssymb,bm}

\usepackage{color}
\usepackage{cite}
\usepackage{amsthm}
\usepackage{amssymb}
\usepackage{varwidth}
\usepackage{algorithmic}
\usepackage{graphicx}
\usepackage{textcomp}
\usepackage{xcolor}
\usepackage[labelformat=simple]{subcaption}
\usepackage{url}
\usepackage{float}
\usepackage{dsfont}
\usepackage{balance}
\usepackage[ruled,vlined]{algorithm2e}
\usepackage{wrapfig}
\usepackage{booktabs}
\usepackage{booktabs}
\usepackage{multirow}
\usepackage[font=small]{caption}
\usepackage[mathcal]{euscript}

\usepackage[caption=false,font=footnotesize]{subfig}
\newcounter{subfigure@save}
\def\BibTeX{{\rm B\kern-.05em{\sc i\kern-.025em b}\kern-.08em
    T\kern-.1667em\lower.7ex\hbox{E}\kern-.125emX}}

\IEEEoverridecommandlockouts
\begin{document}


\title{Hierarchical Federated Learning for Unsupervised Waveform Classification over Tactical MANETs  \thanks{This work was supported by The Boeing Company under agreement no. 2022-540. Correspondence: (cthorn14@vt.edu)}
}

\author{
    \IEEEauthorblockN{
         Charles E. Thornton and
         Daniel J. Jakubisin
    }
    \IEEEauthorblockA{National Security Institute and Bradley Dept. of ECE\\ Virginia Tech \\ Blacksburg, VA, USA 24061\\ 
    }
}

\maketitle

\begin{abstract}
Distributed radio frequency sensing in contested tactical environments demands collaborative learning across mobile nodes. In ad-hoc networks, learning must occur without persistent backhaul, ground truth labels, or reliable communication links. Traditional federated learning approaches assume either ideal link conditions or supervised training objectives, neither of which holds in practice for deployed MANET platforms. This paper presents a hierarchical federated learning framework for unsupervised waveform classification over tactical MANETs subject to Rayleigh fading, random waypoint mobility, and multi-hop routing loss. Each node trains a local denoising convolutional autoencoder on raw IQ observations without label exchange, learning compact representations through a self-supervised reconstruction objective. A two-stage aggregation protocol elects connectivity-based relay aggregators consistent with OLSR multipoint relay selection, compressing cluster-level model updates before forwarding to a mobile server proxy. Across five random seeds, in-network aggregation reduces attempted transmission bits by 16\% on average relative to relay-forward federated averaging at comparable classification performance. Despite mean per-round update drop rates of 21\% (hierarchical) and 34\% (flat), hierarchical MANET FL attains the highest mean unsupervised representation quality of the three federated conditions tested while also showing substantially lower run-to-run variance than either flat MANET FedAvg or ideal FedAvg; flat routing itself shows no consistent benefit or penalty relative to ideal FedAvg. Performance is assessed using KMeans normalized mutual information and linear probe accuracy on the learned latent embeddings.
\end{abstract}

\section{Introduction}
Mobile ad-hoc networks (MANETs) present a fundamental challenge for distributed machine learning. Namely, the environments in which collaborative inference is most valuable are those in which the infrastructure assumptions underpinning conventional learning techniques break down most severely. In MANETs, nodes are mobile, communication links are intermittent and time-varying, and the electromagnetic environment is dynamic and potentially contested. In such settings, the standard approach of routing observations to a central server for batch training is untenable due to a lack of persistent backhaul and the risk of transmitting raw sensor data. Distributed learning must therefore occur at the edge, under the same link conditions that make centralized approaches impractical.

Sidelink communication, standardized in 3GPP Release 16 and extended in Release 17 for operation outside network coverage \cite{harounabadi2021v2x}, enables direct device-to-device transmission without infrastructure support, making it a natural candidate for tactical MANET deployments. Military adaptations of sidelink-style architectures, including waveforms such as Soldier Radio Waveform (SRW) and the emerging family of JADC2-aligned tactical data links, provide multi-hop routing and peer discovery in denied environments \cite{ThorntonMILCOM}. Cognitive radio and dynamic spectrum access techniques have further extended these capabilities by enabling nodes to sense and adapt to the local electromagnetic environment \cite{martone2021closing}. However, the machine learning components of these systems are typically trained offline on curated datasets and deployed as fixed classifiers, with no mechanism for nodes to collaboratively refine their models in response to the evolving signal environment they collectively observe. The result is a system that is adaptive at the waveform layer but static at the inference layer, a gap that becomes increasingly problematic as adversaries field agile, low-probability-of-intercept emitters that fall outside the support of any fixed training distribution.

Federated learning (FL) \cite{mcmahan2017} has emerged as a promising framework for distributed model training, replacing raw data exchange with the aggregation of locally computed model updates. FedAvg \cite{mcmahan2017} demonstrated that a global model can be learned from non-IID data distributed across many clients by averaging local gradient updates. Subsequent work has extended this foundation in several directions. Convergence guarantees under heterogeneous data distributions have been established in \cite{li2020convergence}, while communication efficiency has been addressed through gradient sparsification \cite{lin2023joint}, quantization \cite{alistarh2017qsgd}, and partial participation \cite{mcmahan2017,lin2023joint}. 

FedProx \cite{li2020fedprox} introduced a proximal regularization term to explicitly bound client drift under strong data heterogeneity, establishing a theoretical connection between non-IID distribution and convergence degradation that motivates our investigation of channel-induced subsampling as a complementary mechanism. Hierarchical FL \cite{liu2020client} introduced a client-edge-cloud aggregation architecture that reduces communication burden on the global server by introducing intermediate aggregation tiers; subsequent work extended this to mobility-aware settings \cite{feng2022mobility}, though both assume infrastructure-based topologies with fixed or statistically modeled edge server reachability rather than the dynamic peer-to-peer link structure of a MANET. 

In wireless communications, FL over fading channels has been studied extensively for infrastructure-based uplinks \cite{amiri2020federated}, with particular attention to over-the-air computation and bandwidth-constrained gradient transmission. The convergence behavior of FL under packet erasure has been analyzed in \cite{packetloss2023}, with results showing that FL can converge even under sustained packet loss---a finding consistent with our empirical observations but derived under tighter infrastructure assumptions.

In the radio frequency machine learning (RFML) domain, supervised federated approaches to automatic modulation recognition (AMR) have been proposed in \cite{wang2021federated, dong2025novel}, demonstrating that FL can reduce data privacy risk relative to centralized AMR while maintaining classification accuracy under class imbalance and varying channel conditions. However, these approaches assume nodes possess labeled training data---an assumption that is untenable when observing unknown or adversarial emitters in a contested spectrum. Unsupervised representation learning from raw RF signals has been explored in centralized settings \cite{o2016unsupervised}, and self-supervised learning has been shown to substantially improve label efficiency for AMR \cite{davaslioglu2022self}. Most recently, FedSSL-AMC \cite{akram2025federated} proposed federated self-supervised pretraining for AMR under non-IID distributions, demonstrating that contrastive representation learning can be effectively federated without label exchange. While FedSSL-AMC addresses label scarcity in a federated setting, it assumes reliable uplink delivery and does not model multi-hop routing loss, topology dynamics, or node mobility. To our knowledge, no existing work simultaneously addresses unsupervised federated waveform classification, hierarchical in-network aggregation, and physically realistic MANET channel modeling. This paper addresses these gaps through the following contributions:

\begin{itemize}
\item A hierarchical FL protocol over tactical MANETs with SNR-aware routing, degree-based relay aggregator election consistent with OLSR MPR selection, and 8-bit quantized delta compression, reducing attempted transmission bits by 16\% on average (13.8--18.4\% across five random seeds) relative to relay-forward federated averaging at comparable classification performance.
\item A self-supervised denoising convolutional autoencoder (CAE) trained entirely on unlabeled IQ observations distributed across heterogeneous MANET nodes. No node holds class labels at any point during federation; a small reference set is applied only post-hoc at the central server to assign names to the discovered clusters.
\item Empirical evidence that hierarchical in-network aggregation attains higher mean unsupervised representation quality than ideal FedAvg while substantially reducing its run-to-run variance, whereas flat relay-forward routing shows neither a consistent benefit nor penalty---indicating that the aggregation topology, not channel loss alone, governs whether moderate MANET link loss is detrimental to federated representation learning.
\item A controlled simulation study comparing hierarchical relay-aggregate FL, flat relay-forward FL, ideal FedAvg, and a centralized upper bound under realistic MANET channel conditions, evaluated using KMeans normalized mutual information and linear probe accuracy on learned latent embeddings without requiring labeled data during distributed training.
\end{itemize}

\section{System Model}
We consider a MANET consisting of $N$ nodes deployed within a bounded square area of length $d_{\mathrm{area}}$ meters. Nodes are heterogeneous in their local signal environments and operate without fixed infrastructure or persistent backhaul. Each node assumes one of three roles within a given federation round: a \textit{client} that trains a local model on observed IQ data; a \textit{relay-aggregator} that receives, compresses, and forwards cluster-level model updates; or a \textit{server proxy} that performs global model aggregation. Role assignment is dynamic and determined at the start of each round based on current link state, as described in Section III. All nodes participate as clients; the relay-aggregator and server proxy roles are additionally assigned to a subset of nodes based on adjacency degree. 

Node positions evolve according to the Random Waypoint (RWP) mobility model \cite{bettstetter2004stochastic}, in which each node independently draws a destination uniformly at random within the deployment area and travels toward it at a speed drawn uniformly from $[v_{\mathrm{min}},v_{\mathrm{max}}]$ m/s.  Upon reaching its destination, a node pauses with probability $p_{\mathrm{pause}}$ before drawing a new waypoint, or immediately draws a new destination otherwise. One federation round corresponds to a duration of $T_{\mathrm{round}}$ seconds  during which each node advances toward its current waypoint by a distance of $v \cdot T_\mathrm{round}$ meters, where $v$ is instantaneous speed. Node positions are updated at the start of each round, yielding a time-varying topology that reflects the mobility dynamics of dismounted or light-vehicle tactical platforms. 

\subsection{Channel Model}
Communication between nodes $i$ and $j$ separated by distance $d_{ij}$ is modeled as a Rayleigh flat-fading channel with path loss. The instantaneous received power is given by
\begin{equation}
    P_{r} = P_{t} \cdot h \cdot d_{ij}^{-\alpha},
\end{equation}
where $P_{t}$ is the transmit power, $h \sim \operatorname{Exp}(1)$ is the Rayleigh fading envelope power (exponentially distributed with unit mean), and $\alpha$ is the path loss exponent. The receiver noise power is given by $N_{0} = kT_{0}B \cdot F$, where $kT_{0}$ is the thermal noise density at 290K, $B$ is the channel bandwidth, and $F$ is the receiver noise figure. 

A transmission is successfully decoded if the instantaneous SNR $\gamma = P_{r}/N_{0}$ exceeds a threshold $\gamma_{\mathrm{th}}$. Since $h \sim \operatorname{Exp}(1)$, the outage probability for a single hop of length $d_{ij}$ has closed form
\begin{equation}
    P_{\mathrm{out}}(d_{ij}) = 1 - \operatorname{exp} \left(-\frac{\gamma_{\mathrm{th}}N_{0}}{P_{t}d_{ij}^{-\alpha}} \right).
\end{equation}

For a multi-hop route of $H$ hops with distances $\{d_{1},...,d_{H}\}$, the end-to-end delivery probability under independent per-hop fading is:
\begin{equation}
    P_{\mathrm{success}} = \prod_{k=1}^{H} \left[ 1-P_{\mathrm{out}}(d_{k}) \right].
\end{equation}

A Bernoulli outcome is drawn from $P_{\mathrm{success}}$ at transmission time to determine whether a given model update is delivered. The channel model assumes narrowband flat fading consistent with a single-carrier tactical radio waveform, and does not model co-channel interference, implicitly assuming a TDMA or FDMA protocol.

\subsection{Adjacency and Routing}
At the start of each round, after node positions are updated, a link $(i,j)$ is admitted to the adjacency graph if the single-hop delivery probability exceeds threshold $p_{\mathrm{admit}}$
\begin{equation}
    A_{ij} = \mathds{1}[1-P_{\mathrm{out}}(d_{ij}) > p_{\mathrm{admit}}],
\end{equation}
where $\mathds{1}[\cdot]$ is the indicator function, and $p_{\mathrm{admit}}$ represents the minimum acceptable long-run link reliability for inclusion in the routing graph. This is consistent with the hello-message based neighbor discovery used in OLSR. Routes are limited to a maximum of $H_{\mathrm{max}}$ hops; packets for which no route exists within this constraint are dropped. A propagation delay model assigns a delay of $\lfloor(H-1)/H_\mathrm{delay}\rfloor$ rounds to updates traversing $H$ hops, capped at $D_{\mathrm{max}}$ rounds, reflecting the latency of store-and-forward multi-hop delivery in a low-duty-cycle tactical environment.

\subsection{Signal and Data Model}
Each node observes IQ frames of length $L$ samples drawn from one of $C$ waveform classes. In this work, we consider $C = 2$ classes: linear frequency modulated (LFM) chirp and orthogonal frequency division multiplexing (OFDM), as representative tactical emitter types corresponding to radar and communications waveforms respectively. Each frame is subjected to RF impairments including carrier frequency offset (CFO), IQ imbalance, timing offset, and AWGN at a per-node SNR drawn from a heterogeneous bucket $[\mathrm{SNR}_{\mathrm{min}}^{i},\mathrm{SNR}_{\mathrm{max}}^{i}]$ assigned at initialization. Waveform class proportions per node are drawn from a Dirichlet distribution with concentration parameter $\alpha_{D}$, yielding non-IID data across nodes, with heterogeneity controlled by $\alpha_{D}$. Smaller values of $\alpha_{D}$ produce more skewed per-node class distributions. By design, no node has access to class labels during federation; labels are used only at evaluation time to assess the quality of learned representations.

\begin{table}[t]
\centering
\caption{System Model Parameters}
\label{tab:params}
\renewcommand{\arraystretch}{1}
\footnotesize
\begin{tabular}{llcc}
\toprule
\textbf{Category} & \textbf{Parameter} & \textbf{Symbol} & \textbf{Value} \\
\midrule
\multirow{3}{*}{Federation}
    & No. nodes        & $N$                        & 16            \\
    & FL rounds              & $R$                        & 80            \\
    & Dirichlet param.    & $\alpha_D$                 & 1.5           \\
\midrule
\multirow{6}{*}{Channel}
    & Area side length       & $d_{\mathrm{area}}$          & 2500 $m$       \\
    & Trans. power         & $P_t$                      & 20.5 dBm      \\
    & Noise figure           & $F$                        & 7.0 dB        \\
    & Path loss exp.     & $\alpha$                   & 4.0           \\
    & SNR thresh.         & $\gamma_{\mathrm{th}}$       & 5.5 dB        \\
    & Link admit thresh.   & $p_{\mathrm{admit}}$         & 0.3           \\
\midrule
\multirow{6}{*}{Mobility}
    & Min node speed         & $v_{\min}$                 & 1.0 m/s       \\
    & Max node speed         & $v_{\max}$                 & 15.0 m/s      \\
    & Pause probability      & $p_{\mathrm{pause}}$         & 0.1           \\
    & Round duration         & $T_{\mathrm{round}}$         & 10 s          \\
    & Max routing hops       & $H_{\max}$                 & 3             \\
    & Delay hop granularity  & $H_{\mathrm{delay}}$       & 2             \\
\midrule
\multirow{4}{*}{Signal}
    & IQ frame length        & $L$                        & 128 samples   \\
    & Waveform classes       & $C$                        & 2 (LFM, OFDM) \\
    & Samples per node       & $S$                        & 512           \\
    & Max aggr. delay  & $D_{\max}$                 & 2 rounds      \\
\bottomrule
\end{tabular}
\end{table}

\section{Proposed Method}
Each node maintains a local denoising CAE that maps noisy IQ observations to a compact latent representation. The encoder $f_{\theta}: \mathbb{R}^{2 \times L} \mapsto \mathbb{R}^{d_{z}}$ consists of three strided 1D convolutional layers with batch normalization and ReLU activations, projecting to a latent vector of dimension $d_{z} = 64$ after a fully connected layer with normalization. The decoder $g_{\theta}: \mathbb{R}^{d_{z}} \mapsto \mathbb{R}^{2 \times L}$ mirrors the encoder with transposed convolutions. IQ frames are represented as two-channel real tensors with in-phase and quadrature components along the channel dimension.

The training objective is the mean squared reconstruction error between the decoder output and the impaired frame prior to noise addition, which is given by
\begin{equation}
    L(\theta) = \mathds{E}_{(\tilde{x},x)} \left[\lVert g_{\theta}(f_{\theta}(\tilde{x})) - x \rVert^{2}_{2} \right],
\end{equation}
where $\tilde{x}$ is the noisy observation and $x$ is the normalized signal prior to noise addition. This denoising objective forces the encoder to capture underlying signal structure rather than noise-specific features. Each node trains for $E$ local epochs using Adam with gradient clipping, producing updated parameters $\theta_{i}^{(r)}$ at the end of round $r$.

\subsection{Quantized Delta Compression}
To reduce uplink communication cost, each node transmits a quantized parameter delta rather than full model weights. At the end of local training, node $i$ computes
\begin{equation}
\Delta_{i}^{r} = \theta_{i}^{(r)} - \theta_{\mathrm{global}}^{(r)}. 
\end{equation}

Each parameter tensor is independently quantized to $8$-bit integers using symmetric per-tensor scalar quantization with scale factor $s_{k} = \operatorname{max} \lvert \Delta[k] \rvert /127$, reducing uplink payload to approximately one quarter of float32 transmission. The scale factor is transmitted along with the quantized tensor for exact dequantization at the receiver. At the relay-aggregator, the cluster average delta is re-quantized before stage 2 transmission, introducing a bounded second quantization error that is accepted in exchange for communication savings of compressed forwarding.

\subsection{Hierarchical Aggregation}
The proposed protocol proceeds in two stages per federation round over the MANET topology of Section II.

\textit{Server Proxy and Relay-Aggregator Election:} After positions are updated and adjacency graph $A^{(r)}$ is computed, the server proxy is elected as the node with the highest adjacency degree. A set of $K$ relay-aggregators is then elected from participating clients by the same criterion: selecting the top-$K$ nodes by degree. Each remaining client is assigned to its geographically nearest relay-aggregator, minimizing expected Stage 1 hop count.

\textit{Stage 1, Client to Relay-Aggregator:} Each client routes its quantized delta to its assigned relay-aggregator via greedy geographic forwarding. A Bernoulli delivery outcome is drawn with success probability $P_{\mathrm{success}}$ as defined in Section II-A. Delivered updates are accumulated at the relay-aggregator. Dropped updates are lost for the round.

\textit{Stage 2, Relay-Aggregator to Server Proxy:} Each relay-aggregator computes a sample-count weighted average of received Stage 1 updates:
\begin{equation}
    \bar{\Delta}_a^{(r)}=\frac{\sum_{i \in \mathcal{G}_a} n_i \hat{\Delta}_i^{(r)}}{\sum_{i \in \mathcal{G}_a} n_i},
\end{equation}
where $\mathcal{G}_{a}$ is the set of clients whose updates were received and $n_{i}$ is the number of local training samples at client $i$. The cluster average is re-quantized and routed to the server proxy subject to a further Bernoulli delivery outcome, with a propagation delay of $\delta_{a} = \lfloor(H_{a} - 1)/H_\mathrm{delay}\rfloor$ rounds capped at $D_{\mathrm{max}}$. A Stage 2 failure silently discards all updates aggregated within that cluster.

\textit{Global Aggregation:} The server proxy applies arriving updates using a FedAvgM-style server optimizer with momentum $\beta$ and server learning rate $\eta_{s}$, reducing to standard FedAvg when $\beta = 0$. Delayed updates from prior rounds are applied upon arrival. Attempted transmission bits are charged per hop regardless of delivery outcome, capturing the full network cost of both successful and failed routing attempts.

\section{Experimental Setup and Baselines}
All experiments are implemented in PyTorch. Local client training within a round is independent across clients and, when multiple GPUs are available, is parallelized across a multiprocessing worker pool with each worker receiving a copy of the current global model state at the start of each round. However, the results reported here were produced on a single-device workstation with clients trained sequentially within each round. The MANET channel, mobility engine, and routing protocol are implemented in a custom Python simulation framework interfaced directly with the FL training loop, enabling tight coupling between physical layer dynamics and federation round timing.

\subsection{Baselines and Experimental Conditions}
Four experimental conditions are evaluated:
\begin{itemize}
    \item \textbf{Hierarchical MANET FL (proposed):} Two-stage relay-aggregate protocol with degree-based aggregator election, Rayleigh fading channel, and RWP mobility as described in Section III.
    \item \textbf{Flat MANET FedAvg (relay-forward):} Identical channel parameters, mobility seed, and client specifications to the proposed method, but each client routes its update directly to the server proxy without intermediate aggregation. All channel and data variables are held common between the two conditions.
    \item \textbf{Ideal FedAvg:} Standard FedAvg \cite{mcmahan2017} with all updates delivered every round, no channel effects, no routing delay, and no drop. Represents the upper bound achievable under perfect communication.
    \item \textbf{Centralized:} A single model trained on a pooled dataset of equivalent size to the total federated data. Represents the upper bound when data locality constraints are removed entirely.
\end{itemize}
To ensure a controlled comparison between the two MANET conditions, client selection schedules are precomputed using a shared random seed and held identical per round within each seed tested. The main results (Section V) are averaged across five random seeds and reported as mean $\pm$ one standard deviation; the higher-outage robustness check in Section V and the extended-class-set experiment referenced in the Conclusion are based on a single representative seed and should be interpreted accordingly.

\begin{figure*}[t]
    \centering
    \includegraphics[scale=0.77]{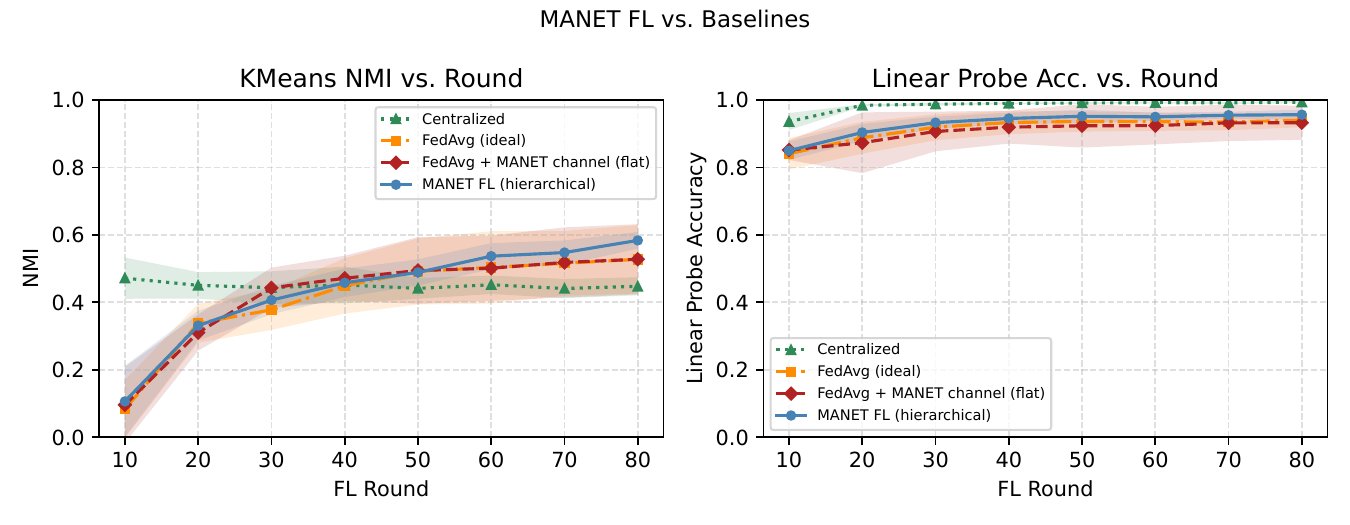}
    \caption{KMeans NMI (left) and linear probe accuracy (right) versus
    FL round, mean $\pm$ 1 standard deviation across five random seeds,
    for hierarchical MANET FL, flat relay-forward MANET FedAvg, ideal
    FedAvg, and the centralized baseline. By round 80, hierarchical
    MANET FL attains the highest mean NMI of the three federated
    conditions (0.58 $\pm$ 0.02), with roughly one quarter the standard deviation of flat MANET FedAvg or ideal FedAvg (both
    0.53 $\pm$ 0.10); the centralized baseline registers the lowest
    mean NMI (0.45 $\pm$ 0.03) despite the highest linear probe
    accuracy throughout (99.3\% $\pm$ 0.3\%). Linear probe accuracy
    shows the same pattern: hierarchical MANET FL reaches the highest mean accuracy among the federated conditions (95.6\% $\pm$ 1.5\%) with the tightest spread, versus 93.2\% $\pm$ 5.1\% (flat) and 94.0\% $\pm$ 2.1\% (ideal).}
    \label{fig:convergence}
\end{figure*}

\subsection{Data Generation}
Training data is generated on-the-fly at each client during local training using a synthetic IQ signal generator. LFM frames are generated with randomized chirp rate and initial frequency drawn uniformly within the 
normalized bandwidth, and a random initial phase. OFDM frames use a 32-subcarrier IFFT with QPSK subcarrier mapping and a cyclic prefix of 25\% of the FFT length, tiled or truncated to $L = 128$ samples. All frames are normalized to unit power before impairment application. RF impairments are applied to each frame independently: CFO drawn from $\mathcal{U}(-5 \times 10^{-4},\, 5 \times 10^{-4})$ normalized frequency, random initial phase offset $\phi_0 \sim \mathcal{U}(0, 2\pi)$, IQ gain imbalance $g \sim \mathcal{U}(-0.003, 
0.003)$, IQ phase imbalance $\vartheta \sim \mathcal{U} (-0.003, 0.003)$, and a small integer timing offset. AWGN is then added at the per-node SNR drawn from the assigned bucket. For the denoising objective, the network input is the impaired noisy frame $\tilde{x}$ and the reconstruction target is the impaired frame prior to noise addition $x$, ensuring the model learns to suppress channel noise while preserving waveform-specific structure introduced by the impairments.

Node class proportions are drawn from a Dirichlet distribution $\mathrm{Dir}(\alpha_D \mathbf{1}_C)$ with $\alpha_D = 1.5$, producing moderate non-IID heterogeneity across nodes. SNR buckets are assigned cyclically across nodes from the set $\{[-5,\,5],\,[5,\,10],\ [5,\,10]\}$~dB, introducing heterogeneity in channel quality in addition to class distribution.

\subsection{Evaluation Metrics and Hyperparameters}
At every $E_{\mathrm{eval}} = 10$ rounds, the global model is evaluated 
on a held-out test set of 800 frames per class generated under the full 
SNR range. Two metrics are computed from the frozen latent embeddings 
$\mathbf{Z} = f_{\theta}(\mathbf{X}_{\mathrm{test}})$:

\textit{KMeans NMI:} KMeans clustering with $K = C = 2$ is applied directly to $\mathbf{Z}$ without label information. Normalized mutual information (NMI) between the cluster assignments and ground truth labels measures the degree to which the unsupervised latent geometry aligns with waveform class structure. NMI ranges from 0 (random alignment) to 1 (perfect alignment) and does not require label access during training or inference.

\textit{Linear Probe Accuracy:} A logistic regression classifier is trained on a 50\% stratified split of the test embeddings using ground truth labels and evaluated on the remaining 50\%. This measures the linear separability of the learned representation --- a lower bound on what a lightweight downstream classifier could achieve, since a nonlinear head could only improve on a linear separator. Importantly, labels are used only at evaluation time and are never transmitted between nodes during federation. 

Local training uses the Adam optimizer with learning rate $\eta = 0.01$, weight decay $10^{-8}$, batch size 64, and $E = 1$ local epoch per round. The server optimizer uses $\beta = 0$  (plain FedAvg aggregation) and $\eta_s = 1.0$. The number of relay-aggregators is $K = 4$, self-updates at aggregators are included, and the maximum aggregation delay is $D_{\max} = 2$ rounds. Full participation is used ($\rho = 1.0$) across all conditions. The centralized baseline trains for $\lfloor R \cdot \rho \rfloor = 80$ epochs to match the effective data exposure of the federated conditions.

\section{Results and Discussion}
\label{sec:results}
\subsection{Convergence Under Channel Impairments}
Fig.~\ref{fig:convergence} shows KMeans NMI and linear probe accuracy versus federation round, averaged across five random seeds, for the three FL conditions and centralized baseline. Despite mean per-round update drop rates of 21\% (hierarchical) and 34\% (flat) under a mean link outage probability of approximately 0.51, both MANET conditions reach linear probe accuracy within roughly 2 percentage points of ideal FedAvg by round 80, demonstrating that the denoising CAE learns discriminative waveform representations under sustained channel impairment.

\begin{figure}
    \centering
    \includegraphics[scale=0.53]{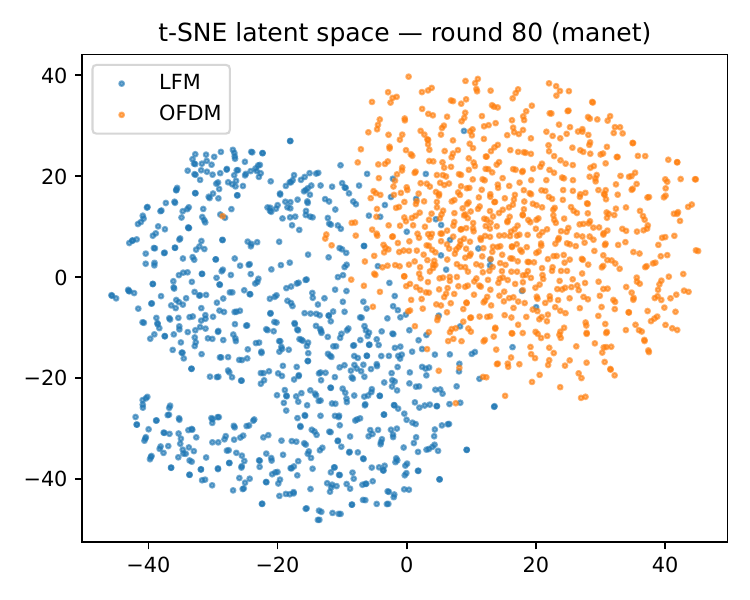}
    \caption{t-SNE projection of latent embeddings learned by hierarchical MANET FL at round 80, for one representative seed. Clear separation between LFM (blue) and OFDM (orange) waveform classes emerges without label supervision during distributed training, with residual overlap concentrated at the class boundary consistent with the linear probe accuracy of 96.25\% observed for this seed.}
    \label{fig:tsne}
\end{figure}

Averaged across five random seeds, hierarchical MANET FL attains the highest mean KMeans NMI of the three federated conditions by round 80 (0.58), compared to 0.53 for both flat MANET FedAvg and ideal FedAvg, with roughly four times lower standard deviation than either (0.02 vs. 0.10). This pattern is not visible from any single seed in isolation: both flat MANET FedAvg and ideal FedAvg range from approximately 0.35 to 0.63 NMI across the five seeds tested, while hierarchical MANET FL stays within a much narrower 0.54--0.61 band in every seed. The centralized baseline registers the lowest mean NMI of all four conditions (0.45) despite the highest linear probe accuracy throughout training (99.3\%) (Fig.~\ref{fig:convergence}, left), suggesting that KMeans NMI on this embedding space does not track linear separability in a simple way and carries meaningful run-to-run uncertainty of its own. Linear probe accuracy shows a consistent pattern: hierarchical MANET FL again attains the highest mean (95.6\%) and lowest variance (std 1.5\%) among the three federated conditions, compared to 93.2\% (std 5.1\%) for flat routing and 94.0\% (std 2.1\%) for ideal FedAvg. Taken together, these results suggest that the hierarchical aggregation topology, rather than channel-induced update loss on its own, is associated with both higher average representation quality and more consistent convergence across independent runs, while flat relay-forward routing shows neither a reliable benefit nor a reliable penalty relative to ideal FedAvg.

\begin{figure}
    \centering
    \includegraphics[width=\columnwidth]{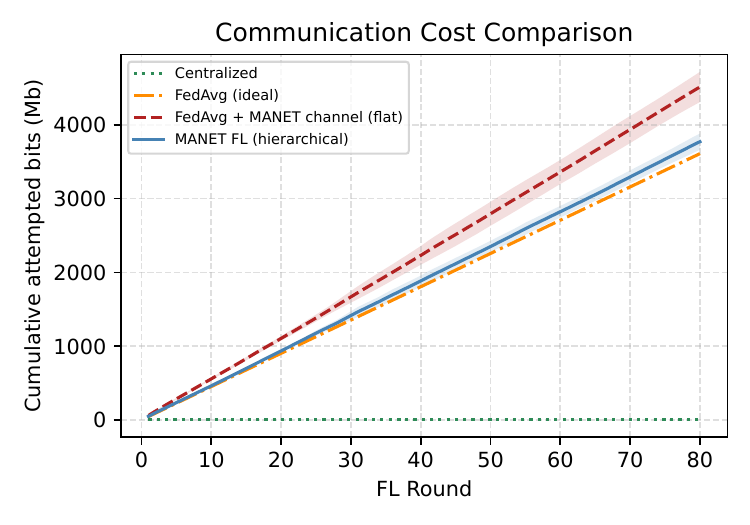}
    \caption{Cumulative attempted bits versus FL round, mean $\pm$ 1 standard deviation across five random seeds. Flat relay-forward MANET FedAvg incurs the highest communication cost due to 16 individual multi-hop transmission attempts per round (4.51 Gb mean cumulative by round 80). Hierarchical MANET FL reduces attempted bits by 16\% on average (13.8--18.4\% range across seeds; 3.77 Gb mean) through shorter Stage 1 hop distances and compressed Stage 2 cluster averaging. Ideal FedAvg accumulates fewer total bits than either MANET condition (3.61 Gb, deterministic since it involves no channel-induced variance) with no multi-hop overhead.}
    \label{fig:commcost}
\end{figure}

Fig. \ref{fig:tsne} shows a t-SNE projection of the learned latent embeddings at round 80, confirming that clear class separation between LFM and OFDM waveforms emerges from the denoising objective without label supervision during federation.

\subsection{Communication Efficiency}
Fig.~\ref{fig:commcost} shows cumulative attempted transmission bits versus FL round. The hierarchical protocol reduces attempted bits by 16\% on average relative to flat relay-forward MANET FedAvg through shorter Stage 1 hop distances and compressed Stage 2 cluster averaging. Ideal FedAvg accumulates fewer bits than either MANET condition as single-hop delivery incurs no multi-hop replication cost.

\subsection{Discussion}
The results suggest that unsupervised waveform classification is viable in a distributed MANET setting without label exchange, and that hierarchical in-network aggregation improves communication efficiency while also attaining higher mean representation quality and substantially lower run-to-run variance than flat relay-forward routing. The centralized upper bound remains ahead on linear probe accuracy by approximately 4--6 percentage points (99.3\% vs. 93.2--95.6\% across the three federated conditions),
reflecting the federated penalty attributable to non-IID data heterogeneity rather than channel impairment; this gap is comparatively stable across seeds for ideal FedAvg and centralized (std $\leq$ 2.1 points) but noticeably wider for flat MANET FedAvg (std 5.1 points), consistent with its higher overall variance. Under the channel conditions tested ($\alpha = 4.0$,
$\gamma_{\mathrm{th}} = 5.5$~dB, mean outage $\approx 0.51$), the hierarchical protocol modestly outperforms flat relay-forward FL on average while being far more consistent across runs. We additionally tested a harsher channel (larger deployment area, mean outage $\approx 0.76$) to probe the cluster-silencing failure mode, in which a Stage~2 drop discards all updates aggregated within a cluster in a single shot. For this single seed, we did not observe the degradation this failure mode might suggest: the hierarchical protocol matched or slightly exceeded flat relay-forward FL on both NMI and linear probe accuracy, with a lower mean drop rate (51\% vs. 56\%) attributable to its shorter Stage~1 hops. This does not rule out the failure mode. A multi-seed study across a range of channel severities would be needed to characterize it precisely. Adaptive switching between relay-aggregate and relay-forward modes based on real-time outage estimation remains a reasonable robustness measure against this architectural exposure and is left for future work.

Two modeling assumptions bound the scope of these results. First, the channel model omits co-channel interference and MAC-layer contention, implicitly assuming orthogonal channel access; under CSMA with hidden terminals the effective delivery probability would degrade further, particularly for Stage 1 traffic concentrated at high-degree aggregators. Second, the Stage 2 cluster-silencing failure mode admits a straightforward architectural remedy: a relay-aggregator may retain its accumulated cluster delta and retry on the subsequent round, or clients may dual-route to the server proxy when their assigned relay's estimated Stage 2 outage exceeds a threshold, recovering flat relay-forward behavior as a limiting case.

\section{Conclusion}
\label{sec:conclusion}

This paper presented a hierarchical federated learning framework for unsupervised waveform classification over tactical MANETs subject to Rayleigh fading, random waypoint mobility, and multi-hop routing loss. A two-stage relay-aggregate protocol with connectivity-based aggregator election reduces attempted transmission bits by 16\% on average relative to relay-forward federated averaging, while a self-supervised denoising CAE enables waveform representation learning without label exchange across nodes. The proposed hierarchical method attains higher mean unsupervised representation quality than both flat relay-forward routing and ideal FedAvg, with substantially lower run-to-run variance, despite a mean per-round update drop rate of 21\% (compared to 34\% for flat routing); flat routing itself shows neither a consistent benefit nor penalty relative to ideal FedAvg. Future work will investigate adaptive protocol switching between relay-aggregate and relay-forward modes based on real-time outage estimation, extension to larger waveform class sets including open-set recognition of unknown emitters, and convergence analysis under the asymmetric staleness introduced by the two-stage delay model. Preliminary centralized experiments extending the class set to four waveform types (LFM, OFDM, QPSK, BPSK) suggest that unsupervised separability in the latent space is class-pair-dependent rather than uniformly degraded by class count: structurally distinct waveforms (LFM, OFDM) remain cleanly separable, while phase-shift-keyed variants with similar per-sample I/Q statistics (BPSK, QPSK) are not reliably distinguished by the reconstruction objective alone. This motivates a discrimination-aware auxiliary loss, rather than reconstruction alone, as a direction for scaling to larger and more fine-grained waveform class sets.

\bibliographystyle{IEEEtran}
\bibliography{sidelinkBib.bib}{}

\end{document}